\documentclass[prb,twocolumn,groupedaddress,showpacs,byrevtex]{revtex4}
\usepackage[]{umlaut}
\usepackage[]{amssymb}
\usepackage[]{graphicx}
\usepackage[]{dcolumn}
\usepackage{bm}

\begin{document}

\bibliographystyle{apsrev}

\title{High Resolution Photoemission Study on Low-$T_{\rm K}$ Ce
  Systems: Kondo Resonance, Crystal Field Structures, and their
  Temperature Dependence}  
\author{D. Ehm}
\author{S. H\"ufner}
\affiliation{Universit\"at des Saarlandes, Fachrichtung 7.2 ---
  Experimentalphysik, 66041 Saarbr\"ucken, Germany} 
\author{F. Reinert}
\email[corresponding author. Email: ]{reinert@physik.uni-wuerzburg.de}
\affiliation{Experimentelle Physik II, Universit\"at W\"urzburg, D-97074 W\"urzburg, Germany}
\author{J. Kroha}
\affiliation{Physikalisches Institut, Universit\"at Bonn, Nussallee 12,
D-53115 Bonn, Germany}
\author{P. W\"olfle}
\affiliation{Institut f\"ur Theorie der Kondensierten Materie,
Universit\"at Karlsruhe, Engesserstr.\ 7, D-76128 Karlsruhe, Germany}
\author{O. Stockert}
\author{C. Geibel}
\affiliation{Max-Planck Institute for Chemical Physics of Solids,
  N\"othnitzer Str. 40, 01187 Dresden, Germany}
\author{H. von L\"ohneysen}
\affiliation{Physikalisches Institut, Universit\"at Karlsruhe,
D-76128 Karlsruhe, Germany}

\date{\today}

\begin{abstract}
We present a high-resolution photoemission study on the strongly
correlated Ce-compounds CeCu$_6$, CeCu$_2$Si$_2$, CeRu$_2$Si$_2$,
CeNi$_2$Ge$_2$, and CeSi$_2$. Using a normalization
procedure based on a division by the Fermi-Dirac distribution we
get access to the spectral density of states up to an energy of $5k_BT$ above
the Fermi energy $E_F$. Thus we can resolve the Kondo resonance and the crystal
field (CF) fine-structure for different temperatures above and around the Kondo temperature
$T_K$. The CF peaks are identified with multiple Kondo resonances within 
the multiorbital Anderson impurity model. Our theoretical $4f$ spectra, calculated from an
extended non-crossing approximation (NCA), describe consistently the observed 
photoemission features and their temperature dependence. By fitting the 
NCA spectra to the experimental data and extrapolating the former to low temperatures,
$T_K$ can be extracted quantitatively. The resulting
values for $T_K$ and the crystal field energies are in excellent
agreement with the results from bulk sensitive measurements,
e.g. inelastic neutron scattering. 
\end{abstract}


\maketitle

\section{Introduction}
\label{intro}

Since the experimental discovery of heavy-fermion (HF) compounds, metallic systems
with rare-earth ($4f$) and actinide ($5f$) elements have been thoroughly investigated both
experimentally and theoretically.\cite{gschneidner,steglich:79} The
term heavy-fermions refers to the 
observation that these systems behave 
as if the conduction electrons had an enormously high effective
mass. The coefficient of the linear term in the low-temperature specific heat,
the Sommerfeld coefficient, can be of the order of 
$1$~JK$^{-2}$mol$^{-1}$, corresponding to an effective electron mass
enhancement of a factor 1000 compared to the free electron mass. In
addition, these $4f$ and $5f$ systems show a variety of anomalous
ground state and unusual low-temperature properties.\cite{stewart84,stewart01}

The single-impurity Anderson model\cite{anderson61} (SIAM)  
embodies the key mechanisms
of the many-body physics in heavy-fermion systems, namely a strong
local Coulomb interaction of the $f$-electrons with the resulting 
local moment formation and a weak hybridization between
conduction electron states and $f$ states, 
which lead to the Kondo effect.\cite{kondo64,schrieffer66}
For cerium systems the theoretical description is significantly 
simplified, because the Ce ions contain at most only one
$4f$-electron (i.e. $4f^1$). Employing variational
methods\cite{gunnarsson83,gunnarsson85}, numerical renormalization 
group (NRG) calculations\cite{bulla07,costi96}  or self-consistent diagrammatic
resummations, like the non-crossing approximation 
(NCA)\cite{bickers87,costi96} 
and the conserving $T$-matrix approximation (CTMA)\cite{kroha97,kroha_review03,
kroha05}, one can calculate
the $f$ spectral density of states (DOS) near the Fermi level within 
the framework of the SIAM. The most important feature in the $4f$ DOS of
Ce-systems is a narrow peak with a maximum just above the Fermi Energy
$E_F$, the {\em Kondo resonance}. The line
width of the Kondo-resonance is given by the low-energy scale of the 
problem, defined by the Kondo-temperature $T_K$. For Ce systems $k_BT_K$ is typically a 
few meV, although the bare model parameters of the SIAM are usually two or 
three orders of magnitude larger. 

Although in the HF compounds the $f$-atoms form an ordered lattice, the
simplest approach is the SIAM, employing non-interacting local impurities, 
which has been very successful in
even a quantitative description of high-resolution photoemission spectra \cite{kondo_reinert01}.
This can be understood as follows: 
The direct orbital overlap between adjacent 
$f$-atoms is negligible, so that the on-site Kondo resonances form 
coherent delocalized states below the lattice coherence temperature only due to
their coupling via the conduction band. Therefore, their dispersion band 
width is essentially given by the Kondo resonance width itself, i.e. by the 
single-impurity Kondo temperature $T_K$. Since photoemission
spectroscopy (PES) measures predominantly the momentum integrated  
f-spectra, these PES spectra are well described by the
SIAM. However, to describe the coherent 
heavy-fermion state with a dispersing narrow band close to the Fermi level,
other models like the renormalized band picture\cite{zwicknagl92} 
or the periodic  Anderson model (PAM)\cite{jarrell95} are required.

PES, in particular angle resolved photoelectron
spectroscopy with excitation energies in the VUV-range (ARUPS), 
is one of the most direct experimental methods to
investigate the electronic structure of solids. For
rare-earth systems, PES allows to study the
occupied part of the $4f$ spectral function with high
accuracy. Unfortunately, the main spectral weight of the Kondo
resonance appears above the Fermi level, where the photoelectron
intensity is suppressed by the Fermi-Dirac distribution (FDD). On the
other hand, {\em inverse} PES (IPES) can in principle
measure the density of states {\em above} $E_F$, but the energy resolution
--- usually considerably larger than 100~meV ---
is not sufficient to investigate the details of the interesting $4f$ spectral 
features, i.e. the Kondo 
resonance, the crystal field structures (CF), and in most cases even not the
spin-orbit splitting. Recently it has been shown\cite{reinert:01:1} on CeCu$_2$Si$_2$
that PES with high energy resolution gives access to the Kondo
resonance (KR). This first and direct observation of the KR was
possible by the
application of a well known normalization procedure \cite{greber:97}
that allows to recover the thermally occupied DOS up to $\sim
5k_BT$ above $E_F$. At higher temperatures this energy range can even
cover the CF structures above the KR. From the normalized PES data one
can determine the Kondo temperature $T_K$ from the KR line width
and the CF splittings of the $J\!=\!5/2$ spin-orbit partner
($\Delta_{CF}$). A more detailed quantitative analysis is possible by 
comparing the experimental $4f$ spectrum with theoretical spectral
functions calculated by the 
 {\em non crossing approximation}
(NCA).\cite{kondo_reinert01} By iteratively fitting the data one can
determine all relevant model parameters of the SIAM,\cite{ehm_sces02} from which
eventually $T_K$ and
$\Delta_{CF}$ can be extracted.

In this paper we present a systematic investigation of the classical
HF systems CeCu$_6$, CeCu$_2$Si$_2$, CeRu$_2$Si$_2$, 
CeNi$_2$Ge$_2$, and CeSi$_2$, which show Kondo temperatures from $5$~K
to about $40$~K as extracted from other experiments. As explained later, we restrict our
investigations to $\gamma$-Ce like materials showing a low hybridization
strength and consequently low Kondo temperatures. We compare the PES results
for the Kondo temperatures and the crystal field energies with values
determined by bulk sensitive measurements.

The manuscript is organized as follows: after a description of the
SIAM, we explain the individual features in the $4f$ density of states and aspects of the 
numerical NCA  calculations (Sec. \ref{sec:theory}). In section \ref{sec:exp_setup}
following section, we describe the experimental setup and the sample
preparation. Section
\ref{sec:results} gives the experimental and theoretical results for
the different HF compounds, followed by a
discussion of the quantitative analysis. Finally, in the appendix we explain
in detail the normalization method and the modeling of the spectra.

\section{Theory of the spectral features}
\label{sec:theory} 

\begin{figure*}[tb]
  \begin{center}
    \includegraphics[width = 14.5cm]{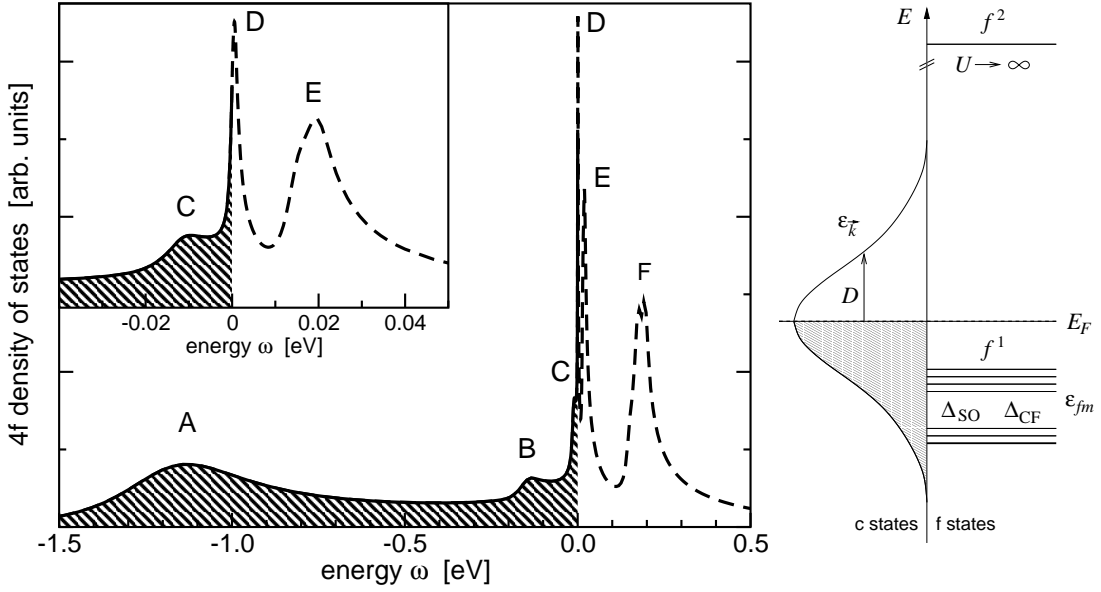}
    \caption[nca_dos]{Left: Theoretical $4f$ spectrum from NCA
      calculations based on the SIAM for $T\!=\!11$ K and the
      model parameters given in the captions of Fig.\ \ref{fig:CeCu2Si2}. The hatched
      region indicates the 'photoemission' region below $E_F$. The inset shows
      the near--$E_F$ region with Kondo resonance ({\sf D}) and crystal
field features ({\sf C,E}). {\sf B} and {\sf F} are the spin-orbit
satellites ($J\!=5/2$), {\sf A} is the ionization peak. For finite
$U$ the two-electron state $f^2$ would appear
      far right of the displayed energy range at 
      $\approx U-|\epsilon_{f}|$. Right: Sketch of the energy level
scheme of the SIAM for the conduction band states
($c$ states) and for the impurity ($f$-states).}
    \label{fig:NCA_DOS}
  \end{center}
\end{figure*}

The SIAM Hamiltonian for Ce in a metallic host reads,
\begin{eqnarray}
H &=& H_{{\rm 0}} + \sum_{m\sigma} \varepsilon _{f m}
f_{m\sigma}^{\dag} f_{m\sigma}^{\phantom{\dag}}
+\sum_{\mathbf{p} m\sigma} 
[V_{\mathbf{p}m} f_{m\sigma}^{\dag} c_{\mathbf{p}\sigma} +h.c.] \nonumber\\
&+&\frac{U}{2} \sum _{(m\sigma)\neq (m'\sigma')} 
f_{m\sigma}^{\dag} f_{m\sigma}^{\phantom{\dag}}\ f_{m'\sigma'}^{\dag} 
f_{m'\sigma'}^{\phantom{\dag}},
\label{hamilton}
\end{eqnarray} 
where 
$H_{{\rm 0}} = \sum_{\mathbf{p}\sigma} 
\varepsilon _{\mathbf{p}} 
c_{\mathbf{p}\sigma}^{\dag} c_{\mathbf{p}\sigma}^{\phantom{\dag}}$
describes the conduction electron band, and $f_{m\sigma}^{\dag}$
creates an electron with spin $\sigma$ in a $4f$-orbital with energy
$\varepsilon _{fm} < E_F$, $m=1,\dots,2S$, $S=7/2$.
The $4f$-orbitals hybridize with the conduction band via the matrix
elements $V_{pm}$. 
For later use we introduce the effective couplings 
$\Gamma _{mm'} = \pi \sum_{\mathbf{p}} V_{m\mathbf{p}}^* V_{\mathbf{p}m'} 
A_{\mathbf{p}\sigma}(0) < 
|\varepsilon_{fm}|$, with $A_{\mathbf{p}\sigma}(\omega)$
the conduction electron spectral function and 
$N(\omega)=\sum _\mathbf{p} A_{\mathbf{p}\sigma}(\omega)$ the conduction
electron density of states per spin, which in the following will be assumed 
to be flat. The Coulomb repulsion $U$ between electrons in any of the local
orbitals is large enough to suppress any double occupancy of the $4f$-shell 
and may be assumed $U\to\infty$ in the following. For the quantum mechanical 
treatment of spin as well as charge fluctuations the $f$-electron operators can then
be represented as $f^{\dag}_{m\sigma}=g^{\dag}_{m\sigma}b$, where the 
bosonic and fermionic auxiliary operators, $b^{\dag}$, $g^{\dag}_{m\sigma}$, 
create the empty and the singly occupied $f$-shell ($m\sigma$), respectively, 
and obey the constraint,
\begin{equation} 
\hat Q = \sum_{m\sigma} g^{\dag}_{m\sigma} g^{\phantom{\dag}}_{m\sigma} +
b^{\dag} b = 1\ .   
\label{constraint}
\end{equation}
The spectrum of this system has generically six 
distinct features as shown in Fig.~\ref{fig:NCA_DOS} ({\sf A}--{\sf F}). 
They can be understood as follows: At low $T$ the occupation of the 
lowest $4f$ level is close to unity, $n_{f1}\lesssim 1$, while
all other $4f$ orbitals are essentially empty,
$n_{fm} \approx 0$, $m=2,\dots,7$. 
Hence, the broad $4f^1$ $\to$ $4f^0$ ionization peak ({\sf A}) 
with a full width at half maximum (FWHM) of $\Gamma \approx \sum_m \Gamma _{1m}$
corresponds to the lowest single-particle
level $\varepsilon _{f1}$. Resonant spin flip scattering of electrons at the 
Fermi energy induces the narrow Kondo resonance ({\sf D}) of width $\sim k_BT_K$,
shifted by $\delta\approx \sin(\pi n_{f1})k_BT_K$ above $E_F$ 
due to level repulsion from the single-particle levels $\varepsilon _{fm}$ (see
Ref.~\onlinecite{kirchner04} for details).
The spin-orbit (SO) and the crystal field (CF) peaks appear in 
pairs ({\sf B}, {\sf F}) and
({\sf C}, {\sf E}), respectively. They arise from virtual transitions from 
the ground state into the (empty) excited 
SO ({\sf F}) and CF ({\sf E}) states and vice versa
({\sf B} and {\sf C}). The positions
of the satellite peak pairs are, therefore, approximately symmetrical about
the central Kondo peak ({\sf D}).
However, while the features above $E_F$ have significant spectral weight,
those below $E_F$ appear merely as weak shoulders. This is because the
transition probabilities carry a detailed balance factor $w = n^i(1-n^f)$,
where $n^i$ ($n^f$) is the occupation number of the $4f$ orbital in the
initial (final) state, i.e. $w$ is large for the excitations {\sf E}, {\sf F},
but small for the transitions {\sf B}, {\sf C}. 
As $n^i$ and $n^f$ are controlled both by quantum and thermal fluctuations 
the CF and SO satellites are $T$-dependent and are signatures of strong
correlations. 

We now analyze the nature of the satellite peaks in more detail.
A straight-forward Schrieffer-Wolff projection onto the subspace of 
the singly occupied $4f$-shell yields an 
$s-d$ exchange model, generalized to multiple local levels, 
\begin{eqnarray}
H_{sd} &=& H_{{\rm 0}} + \sum_{m\sigma} \varepsilon _{f m}
g_{m\sigma}^{\dag} g_{m\sigma}^{\phantom{\dag}} \label{hamilton_sd} \\
&+&\sum_{\mathbf{p} \mathbf{p}' m m'} 
\sum _{\stackrel{\sigma_1\sigma_2\sigma_3\sigma_4}
                {\sigma_1+\sigma_3=\sigma_2+\sigma_4}} J_{mm'} 
c_{\mathbf{p}\sigma_1}^{\dag} c_{\mathbf{p}'\sigma _2}^{\phantom{\dag}} 
g_{m\sigma_3}^{\dag} g_{m'\sigma_4}^{\phantom{\dag}} \ ,
\nonumber 
\end{eqnarray}
subject to the constraint (\ref{constraint}) with vanishing boson number.
The effective spin exchange couplings, including level renormalizations to
second order in $V_{\mathbf{p}m}$ (for non-degenerate levels), 
are obtained as,\cite{kondo_reinert01} 
\begin{equation}
J_{mm'}= \frac{\sum_{\mathbf{p}} V_{m\mathbf{p}}^* V_{\mathbf{p}m'}} 
{\Bigl| \varepsilon _{f1} +
\sum_{n > 1} \frac{|\sum _{\mathbf{p}} V_{\mathbf{p}n}|^2}{\varepsilon_{fn}-
\varepsilon_{f1} }\Bigr| }.
\label{spincoupling}
\end{equation}
Due to the anti-commutative operator structure of the spin coupling term in the 
Hamiltonian (\ref{hamilton_sd}) the conduction electron-local spin T-matrix 
acquires in second order perturbation theory in the $J$s logarithmic 
divergences at the transition energies between the local levels,
$\omega = \varepsilon_{fn}-\varepsilon_{fm}$,
\begin{eqnarray}
T_{mm'}(\omega,T) &=& - N(0) \sum _n J_{mn}J_{nm'}  \\
&\times&
\ln \Bigl| \frac{(\omega +\varepsilon_m - \varepsilon_n)^2+\pi T^2}{D^2}
\Bigr| \ .
\nonumber \label{divergence}
\end{eqnarray}
This expression is derived in a straight-forward way using the bare auxiliary 
fermion propagator $G_{m\sigma}^{(0)}(\nu)=(\nu-\lambda-\varepsilon_m )^{-1}$,
where the parameter $\lambda$ is taken to infinity to project onto the 
constrained Hilbert space of single occupancy of the $4f$-shell 
(see e.g. Ref.~\onlinecite{costi96}). 
$m$ and $m'$ denote which of the $4f$-orbitals is 
occupied in the incoming and in the outgoing channel, respectively, $\omega$ 
is the energy of the scattering conduction electron, and the incoming local
particle is assumed to be at the eigenenergy of the initial 
orbital, $\varepsilon_{fm}$
(on-shell).
Due to the detailed balance 
factors mentioned above, the divergencies in Eq.~(\ref{divergence}) 
give a significant contribution to the $4f$ spectral density only, 
if at least one of the levels $m$ or $n$ is the local ground state. 
The logarithmic energy and temperature dependence demonstrates that the
SO and CF satellite peaks are, in fact, Kondo resonances, i.e.
induced by quantum spin flip processes, however shifted by the 
excitation energies
$\omega=\pm (\varepsilon_{fm}-\varepsilon_{f1})$ with respect to the 
central Kondo peak near $\omega=0$. Despite these multiple Kondo peaks
a single Kondo temperature $T_K$ is defined as the crossover scale
below which the collective spin singlet ground state is formed. 
Since at temperatures below the SO and CF splitting energies only the
$4f$ ground state level $\varepsilon_{f1}$ is significantly occupied,
it may be estimated as,\cite{kondo_reinert01}
\begin{equation}
T_K \approx \sqrt{2 J_{00} E_F}\;  {\rm e}^{-{1}/{(2N(0)J_{00})}}
\end{equation}
and is, hence, given by the width of the {\em central} Kondo peak
near the Fermi energy ({\sf D} in Fig.~\ref{fig:NCA_DOS}). The
phonon-induced broadening of the Kondo resonance and its
satellites is not considered here, because the coupling of the phonons to the 
spin excitations is small.
A detailed analysis of the 
spectral weights and widths of the multiple resonances will be given 
elsewhere, see also Ref.~\onlinecite{kroha03}.

The self-consistent NCA equations are formulated in terms of the 
auxiliary fermion and boson propagators $G_{f}$, $G_{b}$,
\begin{eqnarray} 
[G_{\sigma}^{-1}]_{mm'}(\nu)&=&
(\nu-\lambda-\varepsilon_{fm})\delta _{mm'}-\Sigma_{g\sigma\, mm'}(\nu)\ \ \\
B^{-1}(\nu)&=&
\nu-\lambda-\varepsilon_{fm})-\Sigma_{b}(\nu)\ ,
\end{eqnarray}
where the auxiliary fermion selfenergy $\Sigma_g$ is a matrix in 
orbital space because of the non-conservation of the orbital degree of
freedom $m$, $\delta _{mm'}$ denotes the Kronecker delta and the 
superscript $-1$ matrix inversion. In NCA the selfenergies
$\Sigma_g$, $\Sigma_b$ and the physical $4f$ spectral function $A_f$ 
are given by,
\begin{eqnarray}
&&\hspace*{-1cm}\Sigma_{g\sigma\, mm'}(\nu)=\Gamma_{mm'}\int 
              d\varepsilon\,
               [1-f(\varepsilon )]
              N_{\sigma}(\varepsilon)B(\nu -\varepsilon )
              \label{sigfNCA}\\
&&\hspace*{-1cm}\Sigma_{b}(\nu )=\sum _{\sigma\, mm'}\Gamma_{mm'}\int 
              d\varepsilon\,
              f(\varepsilon )N_{\sigma}(\varepsilon)
              G_{\sigma\, mm'}(\nu +\varepsilon )
              \label{sigbNCA}\\
&&\hspace*{-1cm}A_{f\sigma\, mm'}(\omega )
         = \int  d\varepsilon\,  {\rm e}^{-\beta\varepsilon}
         [ A_{g\sigma\, mm'}(\omega +\varepsilon )A_{b}(\varepsilon )
          \nonumber\\
         &\ &\hspace*{2.8cm}+A_{g\sigma\, mm'}(\varepsilon )
                   A_{b}(\varepsilon -\omega ) ]
              \label{gdNCA}                             
\end{eqnarray}
where all propagators are understood as the retarded ones, and 
the spectral functions are $A_g(\nu)=-{\rm Im}G(\nu)/\pi$, etc.
See Refs.~\onlinecite{costi96},~\onlinecite{kroha_review03} for 
the exact projection onto the physical Hilbert space of no 
multiple $4f$-shell occupation and for an efficient numerical 
evaluation of the NCA equations.

\begin{figure}[tb]
  \begin{center}
 \includegraphics[width = 8.2cm]{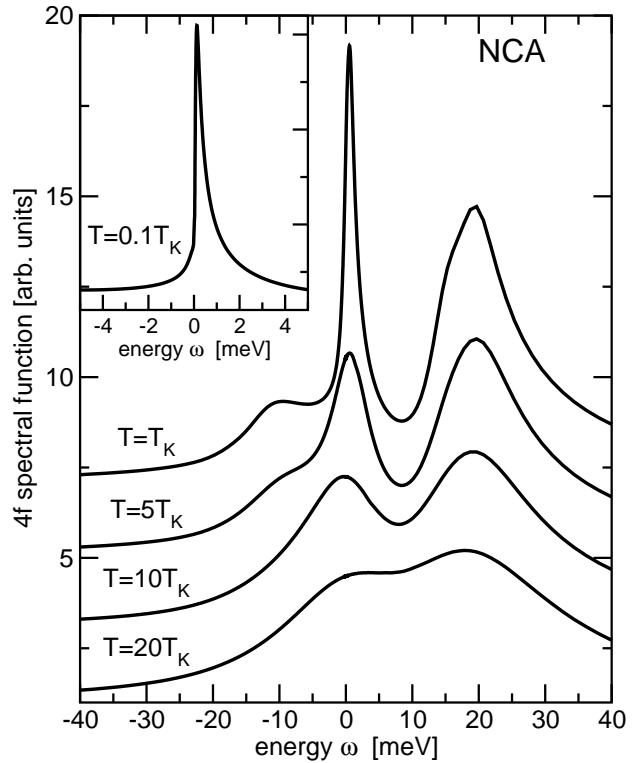}
    \caption[]{Theoretical $4f$ spectrum from NCA
      calculations based on SIAM for different temperatures. The
displayed energy range covers the Kondo resonance close to zero and the crystal
field structures at approx. $-10$~meV and $+20$~meV. The
inset shows the Kondo resonance at low temperatures, i.e. at
      $T=0.1T_K$ (unitarity limit).}
    \label{fig:NCA_Tseries}
  \end{center}
\end{figure}

The NCA is known to correctly include the logarithmic perturbative
corrections in the absence of a magnetic field\cite{kirchner04,kirchner02}
and to reproduce the correct Kondo peak width,\cite{costi96} as long as 
the temperature is not too far below $T_K$. It may also be shown to 
incorporate the above-mentioned detailed balance factors and, hence,
describes well the spectral features discussed above  
(Fig.\ \ref{fig:NCA_Tseries}). 

In order to estimate $T_K$ of the experimental results by fitting of the 
NCA spectra, one calculates the
respective NCA spectra at $T=0.1 T_K\approx 1$~K, i.e. close to the
unitarity limit. $T_K$ can then be determined from the peak width of
this theoretical low-$T$ spectrum (see inset of
Fig.~\ref{fig:NCA_Tseries}).
As already seen in Fig.~\ref{fig:NCA_DOS} at low $T$, i.e. $T\approx
T_K$, the central KR appears as a narrow line, well separated from the CF
satellites above and below the Fermi level. Towards higher
temperatures, the line width of the KR increases, while the maximum
intensity becomes smaller and the distinction of the KR and the CF
features is successively smeared
out. At $T\gg T_K$ the KR disappears as separate peak and an enhanced
$4f$ density of states (DOS) near $E_F$ is rather due to the CF
excitations. Thus, the persistence of an enhanced DOS at the Fermi
level even at high $T$ in comparison to $T_K$ is naturally explained
within the SIAM in combination with CF excited states and their 
logarithmically wide extension towards high $T$.

\section{Experimental Setup}
\label{sec:exp_setup}

The PES experiments have been performed with a SCIENTA SES~200
analyzer in combination with a monochromatized GAMMADATA VUV lamp at photon energies of
$h\nu\!=\!21.23$~eV (He~I$_{\alpha}$) and $h\nu\!=\!40.8$~eV (He~II$_{\alpha}$).
The base pressure of the UHV system was below $5\times 10^{-11}$~mbar,
increasing during the measurements --- due to the He leakage from the discharge lamp --- to
$\lesssim 1\times 10^{-9}$~mbar. The samples
could be cooled down to approximately $T\!=\!4$~K on the manipulator in
the spectrometer chamber. For the presented data analysis the
calibration of the spectrometer is very crucial, in particular the
sample (surface) temperature, the energy resolution, and the position
of the Fermi level. For this purpose we have repeatedly performed
low-temperature reference measurements on polycrystalline Ag and on poly-crystalline Nb in the
superconducting state. By an analysis of these data one can
independently determine the required parameters.\cite{bcs_reinert00,reinert_icess8} 
For the measurements presented here the
energy resolution of the spectrometer was chosen to $5.4$~meV as a
compromise between intensity --- in particular when using
He~II$_{\alpha}$ radiation --- and energy resolution. More about the
spectrometer can be found in Ref.~\onlinecite{cuagau_reinert01}

\subsection{Sample preparation} 
The five different Ce compounds investigated in this work belong to
the class of HF compounds which are characterized by large specific
heat coefficients $\gamma_0$. Whereas CeCu$_6$ displays with
$\gamma_0=1.6$~JK$^{-2}$mol$^{-1}$ the largest value\cite{stewart:84}
among the investigated compounds,  the other systems range from
$\gamma_0=1.1$~JK$^{-2}$mol$^{-1}$ for
CeCu$_2$Si$_2$\cite{lieke:82,steglich:84} down to
$\gamma_0=0.104$~JK$^{-2}$mol$^{-1}$ for CeSi$_2$.\cite{yashima:82:2} 
(CeRu$_2$Si$_2$\cite{steglich:85} and CeNi$_2$Ge$_2$\cite{knopp:88}:
$\gamma_0=350$~mJK$^{-2}$mol$^{-1}$).
None of these compounds shows any magnetic order down to temperatures of
a few
mK.\cite{steglich:79,knopp:88,regnault:87,zemirli:85,loehneysen:93,mori:84,labroo:90,besnus:87}
In the case of CeCu$_2$Si$_2$ and CeNi$_2$Ge$_2$ there is a superconducting phase
transition at $T_c\!=\!0.5$~K\cite{steglich:79} and $0.1$~K\cite{gegenwart:99}, respectively.

The single-crystalline CeCu$_6$ samples were grown by Czochralski
technique in a high-purity argon atmosphere using a tungsten
crucible. In contrast to this the poly-crystals CeSi$_2$,
CeCu$_2$Si$_2$, CeRu$_2$Si$_2$, and CeNi$_2$Ge$_2$ were produced by
first melting the respective stoichiometric ingredients also under high-purity argon
atmosphere and then tempering in a dynamic vacuum. The purity of
the used elements amounts to $99.99$~\% at least. The temperature and
the 
duration of the temper process varies from system to system between
$800$--$1200^{\circ}$C and $50$--$120$~hours. All these compounds
were characterized as single phase compounds by the use of x-ray
diffraction technique, that also yields the crystal structure and
lattice parameters. Whereas all the ternary compounds crystallize in
the tetragonal ThCr$_2$Si$_2$-structure
\cite{jarlborg:83,gupta:83,fukuhara:97} with lattice constants
$a\!=\!4.10$~{\AA}, $4.20$~{\AA}, and $4.15$~{\AA}, and $c\!=\!9.93$~{\AA},
$9.80$~{\AA}, and $9.85$~{\AA} for CeCu$_2$Si$_2$, CeRu$_2$Si$_2$, and
CeNi$_2$Ge$_2$, respectively, CeSi$_2$ crystallizes in the tetragonal
$\alpha$-ThSi$_2$-structure \cite{yashima:82:0} with $a\!=\!4.19$~{\AA}
and $c\!=\!13.91$~{\AA}. The crystal structure of CeCu$_6$ undergoes a
change from orthorhombic symmetry at temperatures above $200$~K to
monoclinic at low temperatures.\cite{asano:86} Because the following
investigations on this compound have all be done in the monoclinic
phase, only the lattice parameters for this structure are given:
$a\!=\!5.08$~{\AA}, $b\!=\!10.12$~{\AA}, $c\!=\!8.07$~{\AA}, and $\beta\!=\!91.36^{\circ}$.

To prepare clean surfaces the samples, cut to a typical size of $2\times3\times5$~mm$^3$ and
equipped with a cleavage post, were
fractured {\em in situ} at low temperatures just before the UPS
measurement. The resulting surfaces were usually coarse grained, even
when the sample was single crystalline. In the latter case, the rough
surface topology ensures an
integration over a sufficiently large effective $k$-space. This is
important, because a strong angular dependence of the $4f$ intensity
was observed.\cite{garnier97R,ehm01} Scraping
the surface with a diamond file did not yield satisfying
results.\cite{yic_reinert98} Because of the high surface reactivity of
the  rare-earth compounds the duration of the measurement was kept
below 12~h. During this period the surface quality was repeatedly checked by a measurement
of the O$\,2p$ photoemission intensity at binding energies of about
6~eV in the valence band spectra.

\section{Results and discussion}
\label{sec:results}

\begin{figure*}[t]
  \begin{center}
    \includegraphics[width = 14.5cm]{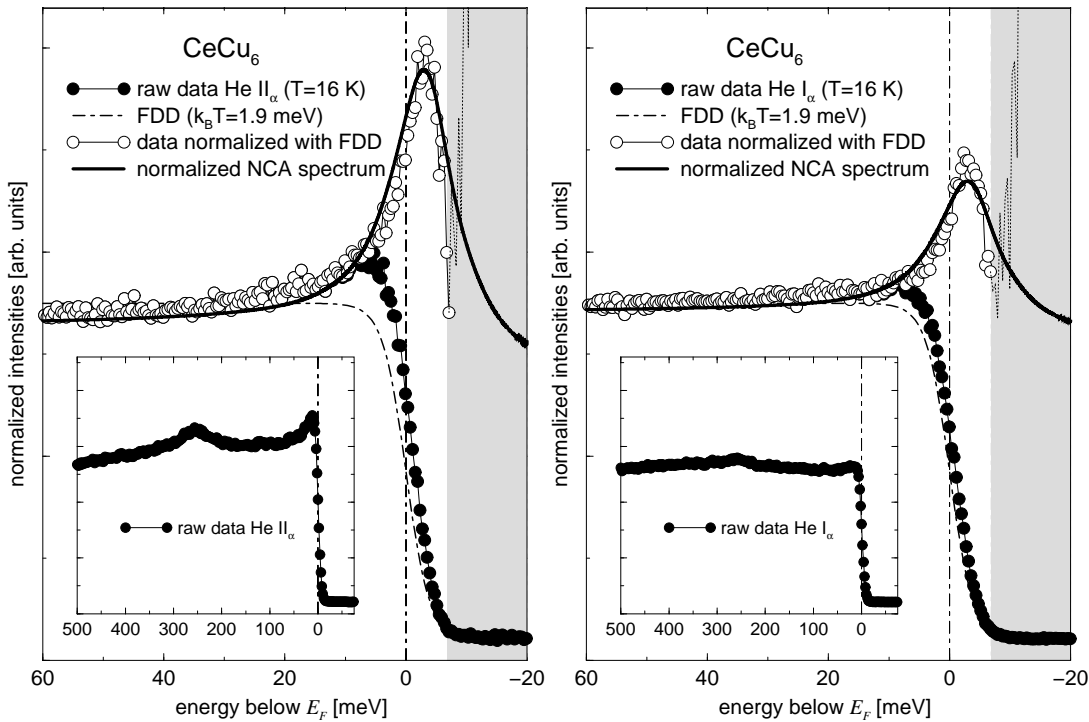}    
    \caption{Photoemission spectrum of CeCu$_6$ (left panel:
He~II$_{\alpha}$ with $h\nu\!=\!40.8$~eV; right panel: He~I$_{\alpha}$ with $h\nu\!=\!21.2$~eV). 
      Filled circles represent the raw data ($\Delta E\!=\!5.4$~meV, $T\!=\!16$~K),
      mainly consisting of the tail of the KR. The SO partner
($J\!=\!7/2$) appears
      at a higher binding energy of $\approx250$~meV given in the inset ($\Delta E\!\approx\!15$~meV). Open circles represent the 
      normalized data using the experimentally broadened FDD,
      the shaded area marks the unreliable spectral range above $5k_BT$. 
      The solid line represents the fitted NCA spectral
      function with 
      $\epsilon_{f}\!=\!-1.05$~eV, $D\!=\!2.8$~eV, CF splittings of the
      $J\!=\!5/2$ sextet $\Delta_{CF}\!=\!7.2/13.9$~meV, SO splitting
      $\Delta_{SO}\!=\!250$~meV, hybridization $V\!=\!116$~meV.}
    \label{fig:CeCu6_lowT}
  \end{center}
\end{figure*}

\begin{figure*}[tb]
  \begin{center}
    \includegraphics[width = 14.5cm]{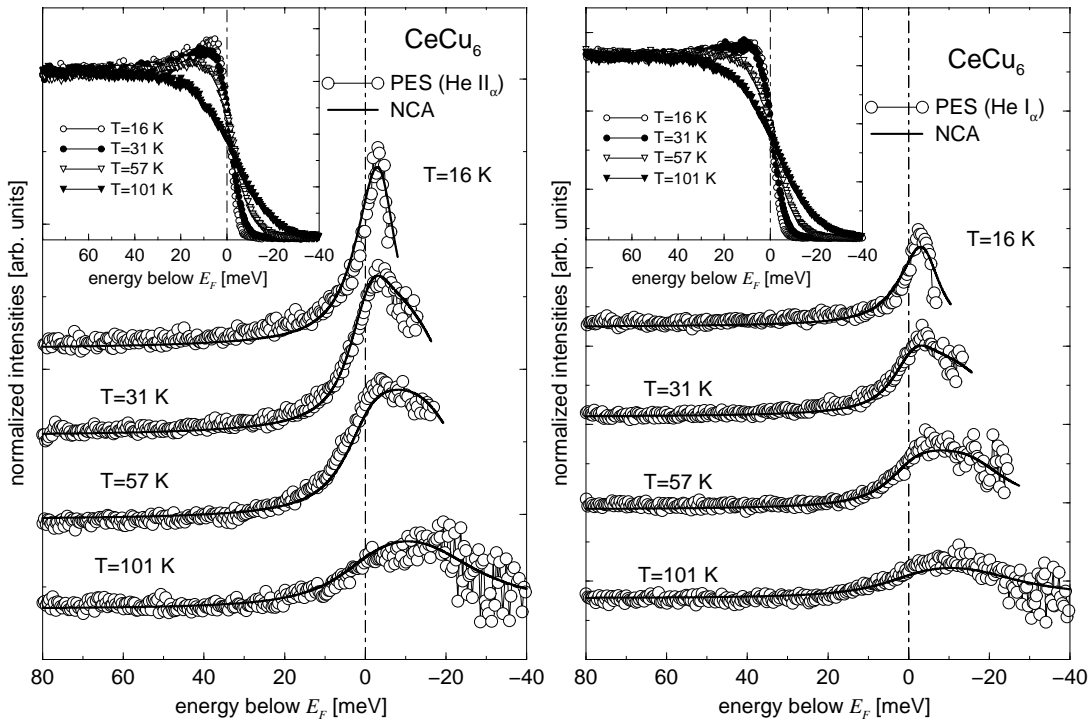}    
    \caption{Temperature dependence of the 
      experimental (PES) and calculated $4f$ spectra of
      CeCu$_6$ after the application of the normalization procedure. 
      The PES data are taken with He~II$_{\alpha}$ (left panel) and
He~I$_{\alpha}$ radiation (right panel).
      The used NCA model parameters are given in the
      caption of Fig.~\ref{fig:CeCu6_lowT}. 
      The insets show the raw spectra, normalized to the same intensity
      at a binding energy of $\approx100$~meV.}
    \label{fig:CeCu6_Tseries}
  \end{center}
\end{figure*}

\begin{figure*}[tb]
  \begin{center}
    \includegraphics[width = 14.5cm]{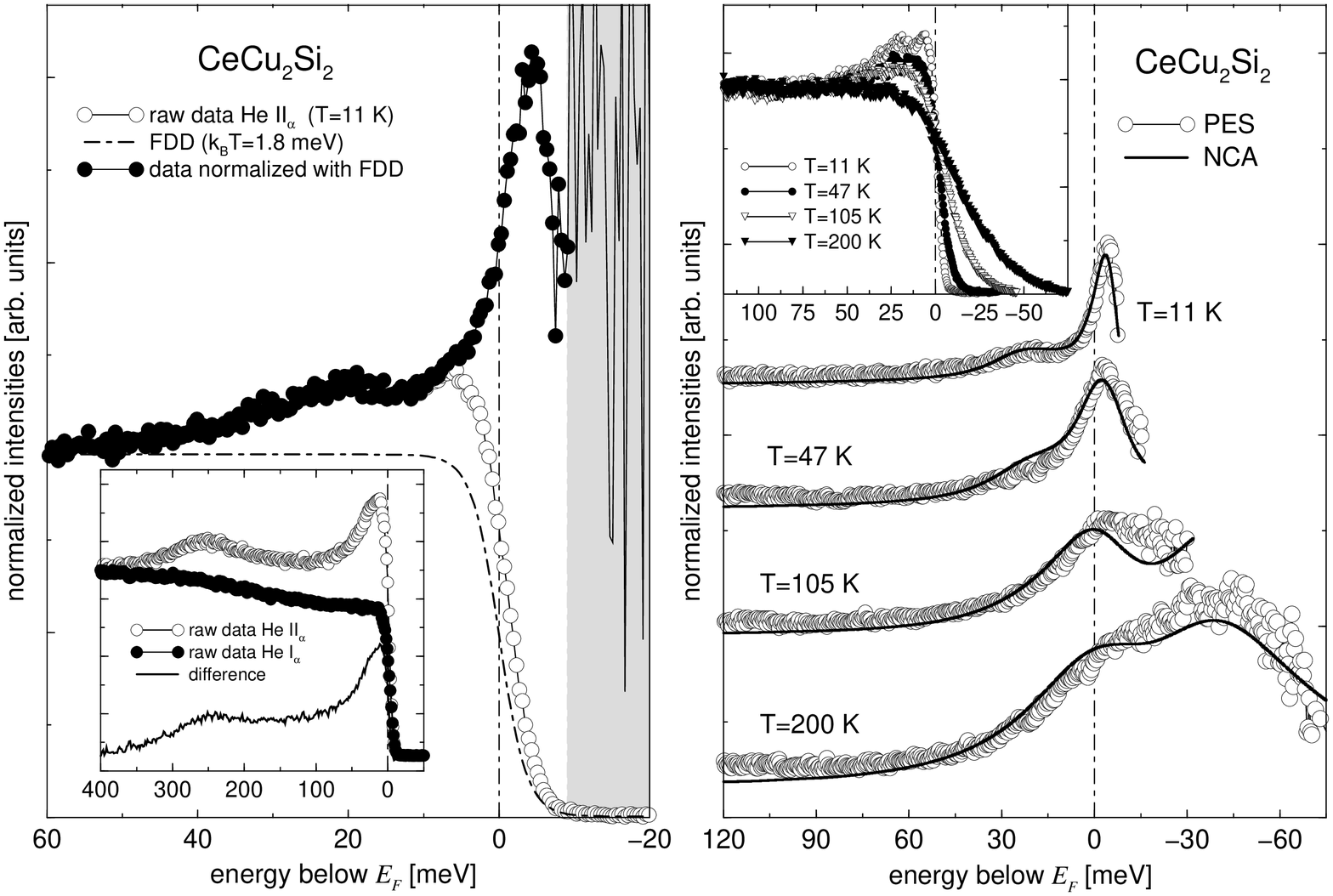}    
    \caption{Photoemission spectra of CeCu$_2$Si$_2$
(He~II$_{\alpha}$, $h\nu\!=\!40.8$~eV). Left panel: normalization at
$T\!=\!11$~K (cf. Fig.~\ref{fig:CeCu6_lowT}); the inset shows an extended energy
      range ($\Delta E\!\approx\!15$~meV) including
      the SO ($J\!=\!7/2$) excitation at $\approx270$~meV below $E_F$
      taken with He~II$_{\alpha}$ radiation (open circles) and with
He~I$_{\alpha}$ (filled circles).
      Right panel: temperature dependence of the experimental (He~II$_{\alpha}$)
and the calculated $4f$ spectral function of
      CeCu$_2$Si$_2$ after the normalization. Model parameters:
      $\epsilon_{f}\!=\!-1.57$~eV, $D\!=\!4.3$~eV,
      $\Delta_{CF}\!=\!32/37$~meV, 
      $\Delta_{SO}\!=\!270$~meV, $V\!=\!200$~meV. The inset
      shows the raw data normalized to the same intensity
      at a binding energy of $\approx100$~meV.}
    \label{fig:CeCu2Si2}
  \end{center}
\end{figure*}

\begin{figure*}[tb]
  \begin{center}
    \includegraphics[width = 14.5cm]{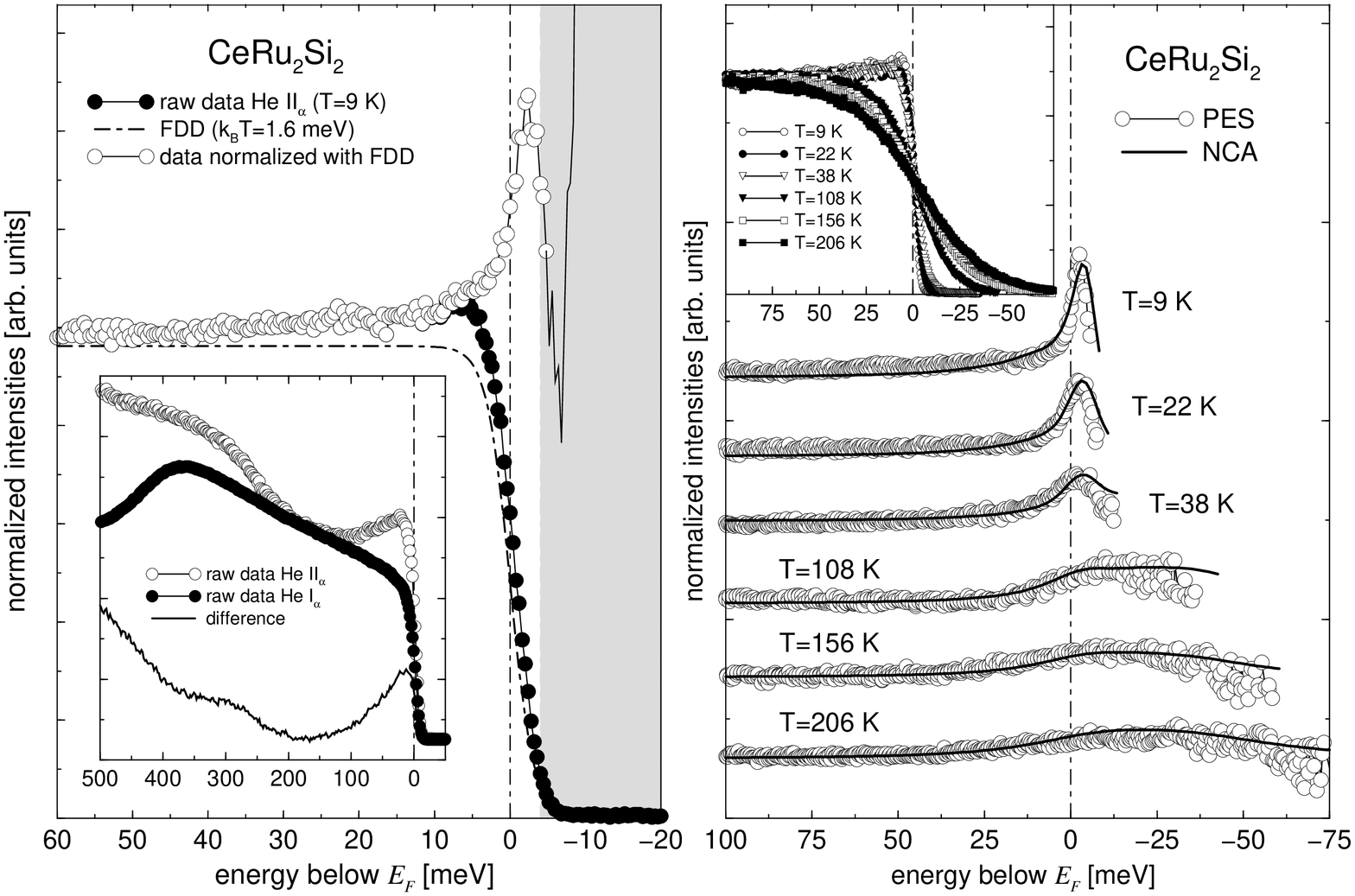}  
  \caption{Photoemission spectra of CeRu$_2$Si$_2$
(He~II$_{\alpha}$, $h\nu\!=\!40.8$~eV). Left panel: normalization at
$T\!=\!11$~K (cf. Fig.~\ref{fig:CeCu6_lowT}); the inset shows an extended energy
      range ($\Delta E\!\approx\!15$~meV) including
      the SO ($J\!=\!7/2$) excitation at $\approx290$~meV below $E_F$
      taken with He~II$_{\alpha}$ radiation (open circles) and with
He~I$_{\alpha}$ (filled circles).
      Right panel: temperature dependence of the experimental (He~II$_{\alpha}$)
and the calculated $4f$ spectral function of
      CeRu$_2$Si$_2$ after the normalization. Model parameters: $\epsilon_{f}\!=\!-1.54$~eV, $D\!=\!4.2$~eV, $\Delta_{CF}\!=\!18/33$~meV,
      $\Delta_{SO}\!=\!295$~meV, $V\!=\!170$~meV. The inset
      shows the raw data normalized to the same intensity
      at a binding energy of $\approx100$~meV.}
    \label{fig:CeRu2Si2}
  \end{center}
\end{figure*}

\begin{figure*}[tb]
  \begin{center}
    \includegraphics[width = 14.5cm]{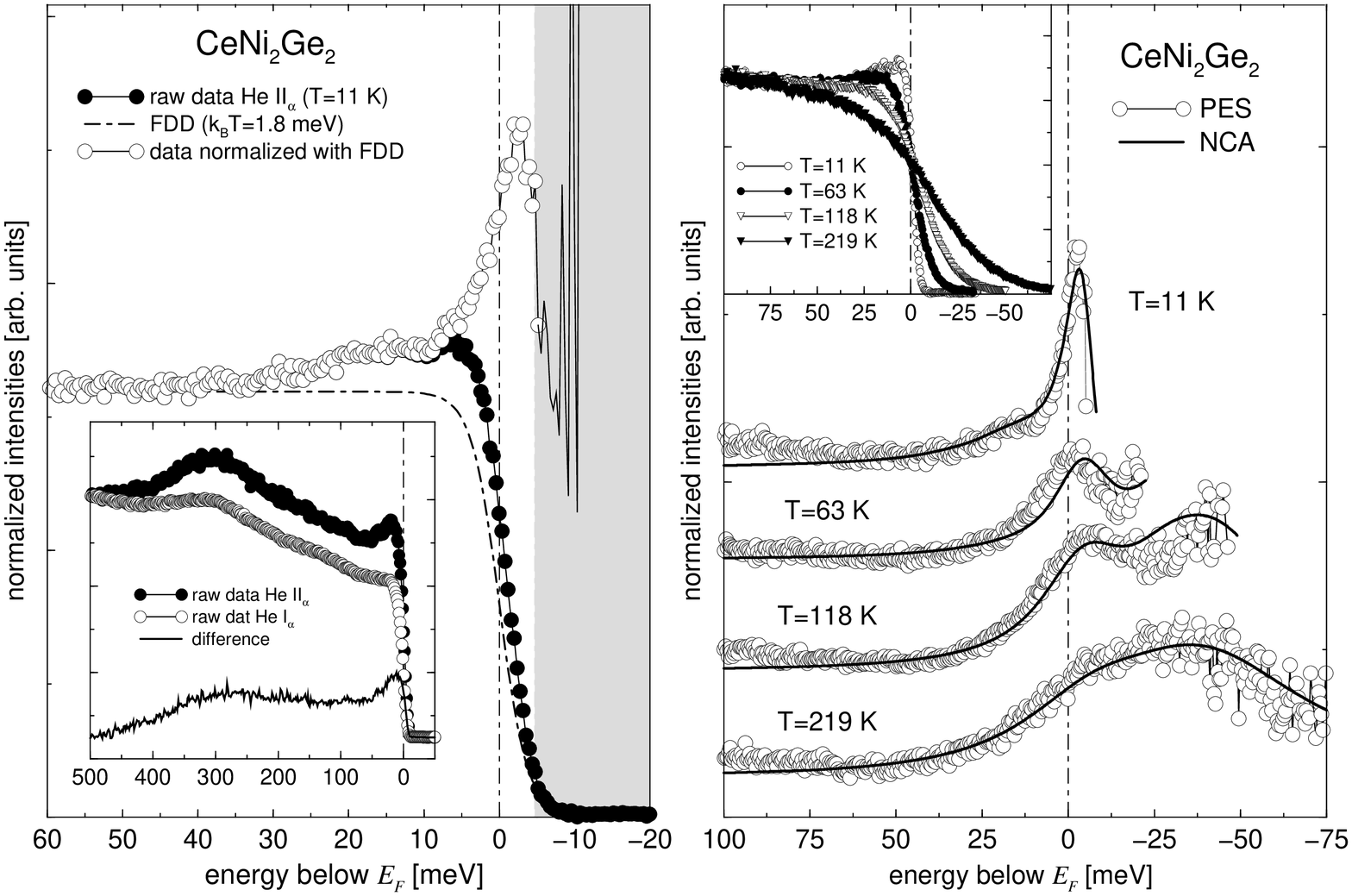}    
    \caption{Photoemission spectra of CeNi$_2$Ge$_2$
(He~II$_{\alpha}$, $h\nu\!=\!40.8$~eV). Left panel: normalization at
$T\!=\!11$~K; the inset shows an extended energy
      range ($\Delta E\!\approx\!15$~meV) including
      the SO ($J\!=\!7/2$) excitation at $\approx300$~meV below $E_F$
      taken with He~II$_{\alpha}$ radiation (open circles) and with
He~I$_{\alpha}$ (filled circles).
      Right panel: temperature dependence of the experimental (He~II$_{\alpha}$)
and the calculated $4f$ spectral function of
      CeNi$_2$Ge$_2$ after the normalization. Model parameters: $\epsilon_{f}\!=\!-1.43$~eV, $D\!=\!3.9$~eV, $\Delta_{CF}\!=\!26/39$~meV,
      $\Delta_{SO}\!=\!275$~meV, $V\!=\!209$~meV. The inset
      shows the raw data normalized to the same intensity
      at a binding energy of $\approx100$~meV.}
    \label{fig:CeNi2Ge2}
  \end{center}
\end{figure*}

\begin{figure*}[tb]
  \begin{center}
    \includegraphics[width = 14.5cm]{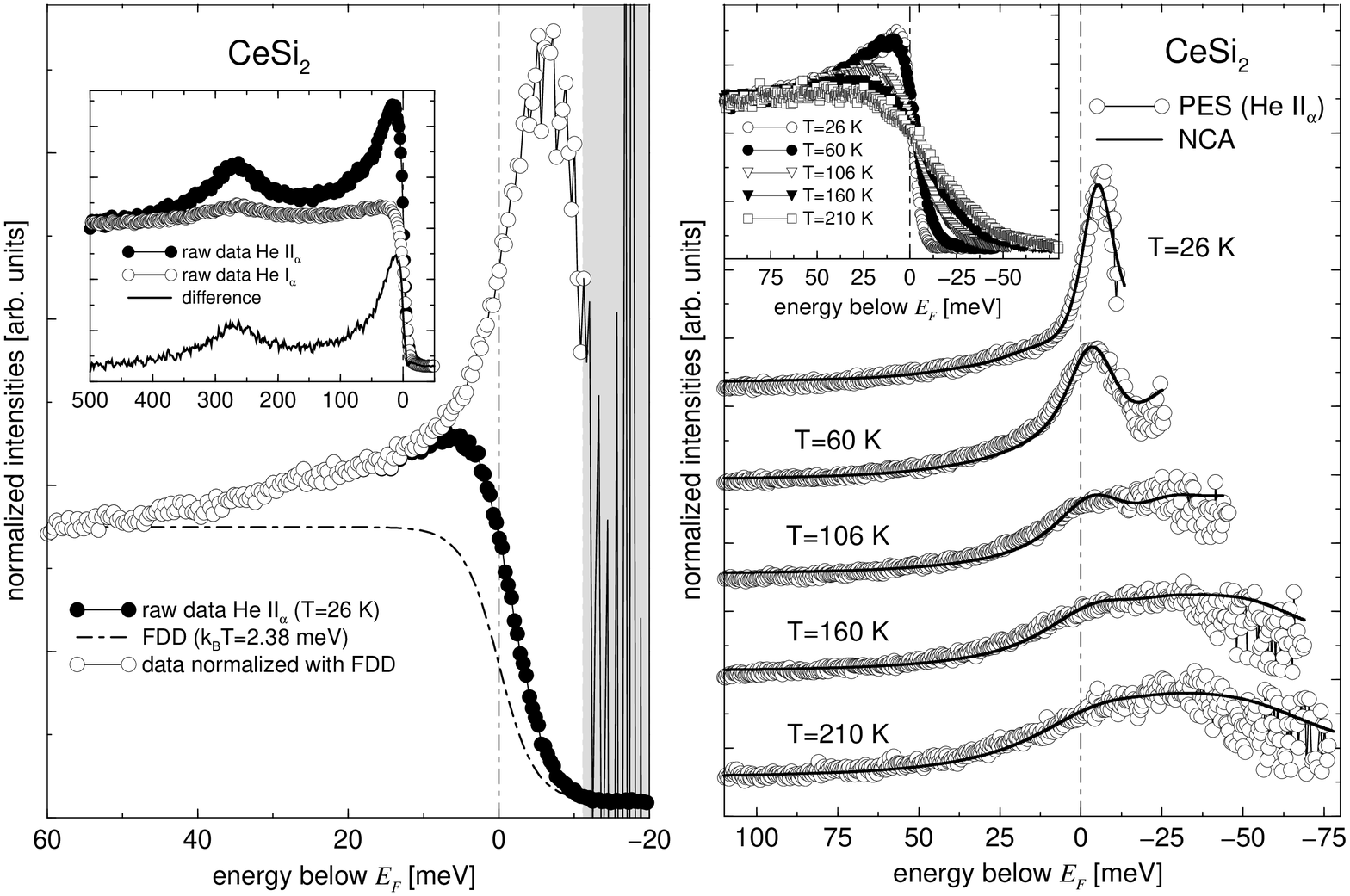}    
    \caption{Photoemission spectra of CeSi$_2$
(He~II$_{\alpha}$, $h\nu\!=\!40.8$~eV). Left panel: normalization at
$T\!=\!11$~K; the inset shows an extended energy
      range ($\Delta E\!\approx\!15$~meV) including
      the SO ($J\!=\!7/2$) excitation at $\approx280$~meV
      taken with He~II$_{\alpha}$ radiation (open circles) and with
He~I$_{\alpha}$ (filled circles).
      Right panel: temperature dependence of the experimental (He~II$_{\alpha}$)
and the calculated $4f$ spectral function of
      CeSi$_2$ after the normalization. Model parameters: $\epsilon_{f}\!=\!-1.35$~eV, $D\!=\!3.7$~eV, $\Delta_{CF}\!=\!25/48$~meV,
      $\Delta_{SO}\!=\!270$~meV, $V\!=\!203$~meV. The inset
      shows the raw data normalized to the same intensity
      at a binding energy of $\approx100$~meV.}
    \label{fig:CeSi2}
  \end{center}
\end{figure*}

\begin{table*}[ht]
  \begin{center}
    \setlength{\tabcolsep}{0.5cm}
    \begin{tabular}[t]{ccccccc}
	\toprule
&\multicolumn{3}{c}{from other experiments}&\multicolumn{3}{c}{from PES}\\
&$T_K$~[K]&\multicolumn{2}{c}{CF
  [meV]}&$T_K$~[K]&\multicolumn{2}{c}{CF [meV]}\\ 
&&$\Delta_{1\rightarrow 2}$&$\Delta_{1\rightarrow 3}$&&$\Delta_{1\rightarrow 2}$&$\Delta_{1\rightarrow 3}$\\ \hline
&&&&&&\\
CeCu$_6$&$5.0\pm 0.5^{(a)}$\cite{rossat-mignod:88}&$7.0^{(a)}$\cite{goremychkin:93}&$13.8^{(a)}$\cite{goremychkin:93}&$4.6$&$7.2$&$13.9$\\ 
\colrule
&&&&&&\\
CeCu$_2$Si$_2$&$4.5^{(b)}$\cite{bredl:84}$-$
$10^{(a)}$\cite{horn:81}&$30^{(a)}$\cite{goremychkin:93}$-$ $36^{(c)}$\cite{cooper:86}&---&$6$&$32$&$37$\\
\colrule
&&&&&&\\
CeRu$_2$Si$_2$& $16^{(a)}$\cite{severing:89:0}&$19^{(d)}$\cite{zwicknagl:92}&$34^{(d)}$\cite{zwicknagl:92}&$16.5$&$18$&$33$\\
\colrule
&&&&&&\\
CeNi$_2$Ge$_2$&$29^{(a)}$\cite{knopp:88}&$(4)^{(a)}$\cite{frost:00}&$34^{(a)}$\cite{frost:00}&$29.5$&$26$&$39$\\
\colrule
&&&&&&\\
CeSi$_2$&$22/41^{(a)}$\cite{galera:89}&$25^{(a)}$\cite{galera:89}&$48^{(a)}$\cite{galera:89}&$35$&$25$&$48$\\
    \botrule
  \end{tabular}

  \caption[]{ \label{tab:TK_CEF}Comparison of the Kondo temperatures $T_K$ and the CF
  energies $\Delta_{CF}$ determined from PES and from other experimental
  methods, namely from 
  $(a)$: INS studies, $(b)$: specific heat measurements, $(c)$: Raman scattering experiments, and  
  $(d)$: theoretical considerations based on specific heat measurements.} 
  \end{center}
\end{table*}

In the following section we present our results on five
HF compounds with Kondo temperatures below $100$~K, namely
CeCu$_6$ ($T_K=5$~K\cite{rossat-mignod:88}), CeCu$_2$Si$_2$
($10$~K\cite{horn:81}), CeRu$_2$Si$_2$ ($20$~K\cite{rossat-mignod:88}),
CeNi$_2$Ge$_2$ ($30$~K\cite{knopp:88}),
($41$~K\cite{galera:89}); the given Kondo temperatures $T_K$ are
determined by inelastic neutron scattering (INS).
Here we restrict our investigations to low-$T_K$ or $\gamma$-Ce like
materials with a low hybridization strength because of two
reasons: First it is a well known fact that these $\gamma$-like
materials only exhibit a small change in hybridization strength by
going from the bulk to the surface that is not the case in
$\alpha$-like Ce compounds.\cite{laubschat:90:1} Recently J. W. Allen
\cite{allen:00} showed that PE spectra of $\gamma$-like Ce compounds
mostly exhibit bulk rather than surface contributions. Second,
the position of the KR in heavy fermion Ce compounds
occurs above $E_F$ at $\delta\approx \sin(\pi n_{f1})k_BT_K$.\cite{kirchner04} 
Because our normalization method 
(described in detail in Appendix~\ref{norm_proc}) only allows to
observe spectral features in the energy range between $E_F$ and
$5k_BT$ above $E_F$, it is a crucial aspect to find the KR
in this energy interval at low temperature.
By using the normalization procedure we are also able to compare the PE
spectra with NCA spectral functions. To get a
quantitative comparison, we have to apply this normalization method to
both the PES data and the NCA results. By the performance of a
conventional $\chi^2$-fitting procedure, we get a suitable parameter set for
the description of the spectral function of Ce systems as described in
Sec.~\ref{sec:theory}.

\subsection{Low-temperature spectra of CeCu$_6$}
The first presented system is the prototype HF
compound CeCu$_6$. Because of the small Kondo-temperature of only
$T_K\!\approx\!5~K$ and the correspondingly weak $4f$ spectral weight at
the Fermi level, there
exist only few valence band photoemission studies on this
compound.\cite{patthey:86,schlapbach:86,chiaia:97,ehm_sces03}
Our high resolution photoemission spectra on CeCu$_6$ are displayed in
Fig.~\ref{fig:CeCu6_lowT} measured at $T\!=\!16$~K with both He~I$_{\alpha}$
(right panel) and He~II$_{\alpha}$ (left panel) radiation. 
The insets show an extended energy range below $E_F$ over $500$~meV
with the weak tail of the Kondo resonance at $E_F$ and the SO feature
at approx. 260~meV corresponding to peak {\sf B} in
Fig.~\ref{fig:NCA_DOS}. Although the photoionization cross section of the Ce~$4f$ states
increases about a factor $3.2$ from an excitation energy of
$h\nu\!=21.23$~eV to $40.8$~eV\cite{yeh_lindau:85} both spectra clearly reveal the $4f$ derived
spectral features over the conduction band background, although the
$4f$ intensity with He~I is significantly
reduced. 

In the blowup of the near-$E_F$ region in Fig.~\ref{fig:CeCu6_lowT},
only the tail of the KR is discernible, with an increasing intensity
towards $E_F$. The spectral information above the Fermi level --- the
maximum intensity of
the Kondo resonance is anticipated slightly above $E_F$ --- is suppressed
by the Fermi-Dirac distribution (FDD) that steeply goes to zero for
small temperatures. The spectral information can partly be restored
when the above mentioned normalization method (see
Appendix~\ref{norm_proc}) is applied to the raw He data.
As demonstrated already for He~II-data on CeCu$_2$Si$_2$ in
Ref.~\onlinecite{kondo_reinert01}, there appears obviously a narrow peak with a full
width at half 
maximum (FWHM) of $\approx6$~meV and a maximum at about $3$~meV
above $E_F$.  The peak can be observed in both the He~II$_{\alpha}$ and in the He~I$_{\alpha}$
spectra, but with a peak-to-background intensity ratio which is about
a factor of 2 larger for He~II. The peak
maximum is sufficiently below the upper limit of $5k_BT\!\approx\!7$~meV for the applicability of
the method $5k_BT\approx 7$~meV, where the noise of the data becomes
enormously magnified due to the exponential
decrease of the FDD. It should be mentioned that the data for
binding energies below $2k_BT$ (below $E_F$) remain unchanged by the
normalization procedure.

In order to analyze the narrow peak feature further, we have performed
NCA calculations with a set of model parameters that is given in
the figure caption of Fig.~\ref{fig:CeCu6_lowT}. The starting points
for the model parameters are
taken from the photoemission spectra at higher energies
($\epsilon_f$, $\Delta_{SO}$) or from other experiments, as
e.g. inelastic neutron scattering ($\Delta_{CF}$), and then reasonably
modified to fit the normalized experimental spectra.  The intensity ratio between $4f$ and conduction
band states, which are assumed to be constant in energy, is set
arbitrarily, the Coulomb energy $U$ is infinite.
The agreement between theory and experiment is striking. The NCA result
is able to describe exactly the experimental line shape and the energetic
position of the narrow peak just above $E_F$ both in the spectra taken with He~II$_{\alpha}$ and He~I$_{\alpha}$
radiation. 

\subsection{Temperature dependence of the CeCu$_6$ $4f$ spectra}
The NCA allows to calculate spectra for finite temperatures. All
physical properties --- and in particular the spectral density of
states\cite{bickers87} --- scale with the Kondo temperature $T_K$. In addition, 
the recoverable energy range (above $E_F$) increases
linearly with $T$, because the thermal broadening of the FDD increases with the sample 
temperature. Therefore we analyzed the near-$E_F$ spectral
structures of CeCu$_6$ for different
temperatures. Fig.~\ref{fig:CeCu6_Tseries} shows the experimental data
plus the NCA result at several temperatures, normalized at the FDD at
the respective temperature (see Appendix~\ref{norm_proc}).
Again we applied the normalization to both the He~II$_{\alpha}$ (left
panel) and the
He~II$_{\alpha}$ (right panel) spectra on CeCu$_6$. The insets give
the raw data, normalized at the same intensity at 
$E_B=100$~meV. 
These data are dominated by the increasing broadening of the FDD
and the vanishing
tail of the Kondo resonance.

As already seen in Fig.~\ref{fig:CeCu6_lowT} at low $T$, the KR
appears in the recovered data at $T\!=16$~K as a narrow line just above $E_F$. Towards
higher temperatures the line width of the KR increases while the
maximum intensity becomes smaller. The spectra of systems which
clearly resolve a crystal field (CF) structure (see below CeCu$_2$Si$_2$
and CeNi$_2$Ge$_2$) show an additional smearing of the crystal field
fine structure. At $T\!=\!101$~K and above, the narrow peak of the KR
has disappeared, and is replaced by a broad structure with a maximum
at $\approx10$~meV above $E_F$, which results in a considerable $4f$ density of
states right at the Fermi level. 

With the same model parameters as for the low-temperature spectrum in
Fig.~\ref{fig:CeCu6_lowT} we have calculated the spectra at the higher
experimental temperatures. The temperature dependence observed in the
experimental data is perfectly reproduced by the NCA result which is
qualitatively described in Fig.~\ref{fig:NCA_Tseries}. As we will
discuss further below, the existence of the crystal field splittings
of the $4f$ levels has a considerable influence on the distribution of
spectral weight near the Fermi level.

\subsection{CeCu$_2$Si$_2$, CeRu$_2$Si$_2$, CeNi$_2$Ge$_2$, CeSi$_2$: Low-$T$ spectra and temperature dependence}

This becomes more clear in the case of the HF-compound CeCu$_2$Si$_2$
with a Kondo temperature of $T_K\approx10$~K and a crystal field
energy of about 30~meV (see references given in
Tab.~\ref{tab:TK_CEF}). At
$E_B\approx20$~meV the raw low-temperature spectra show a
clearly distinct crystal-field structure in the $4f_{5/2}$
levels.\cite{goremychkin:93} A similar CF structure
could also be observed in CeB$_6$\cite{souma01} and is also
visible in the CeNi$_2$Ge$_2$ spectra below. 
In addition to this fine structure, the high-temperature data show a 
large shift of the broad peak maximum to approx. 45~meV above $E_F$,
even a two-fold fine structure in the high-temperature peak can be
resolved. Both features, the CF satellite below $E_F$ in the
low-temperature data and the broad split peak at high temperatures are
well described by the NCA calculation {\em including} the crystal
field resolved degeneracy of the $4f$ levels. Without this CF
splitting, there is only the spin-orbit splitting on the scale of
$\Delta_{SO}=250$~meV, which does not contribute to the investigated
energy range, and the observed temperature dependence of the spectra
can not be described in the SIAM quantitatively. The high spectral
intensity at $E_F$ at temperatures $T\gg T_K$ is an immediate result of
the existence of the crystal field structures.

We performed analogous investigations on the HF compounds CeRu$_2$Si$_2$,
CeNi$_2$Ge$_2$, and CeSi$_2$ which exhibit $T_K$'s in ascending order
from $16$ to $40$~K. The results of our PES investigations are shown in
Figs.~\ref{fig:CeRu2Si2}--\ref{fig:CeSi2}. In each figure the left
panel shows the analysis of the low-temperature data and the right
panel gives the temperature dependence of the near-$E_F$ spectral
features. In the following we
restrict ourselves to the He~II$_{\alpha}$ 
data because all $4f$ states are much more
pronounced than at low photon energies (He~I). 
Especially in the case of the ternary Ce compounds displayed here, the
intensity of the $4f$ spectral features below $E_F$ is reduced dramatically, in
contrast to CeCu$_6$ for which the $4f$ features could still be observed
in the He~I data. Even by irradiation with He~II$_{\alpha}$ the
near-$E_F$ spectra of CeRu$_2$Si$_2$, CeNi$_2$Ge$_2$, and CeSi$_2$
show only a weak $4f$ fine structure. 

A comparison of the He~I and He~II spectra is given in the insets of
Figs.~\ref{fig:CeCu2Si2}--\ref{fig:CeSi2}. Taking the difference of
the two spectra\cite{garnier97} is a useful method to extract the
$4f$ contribution from other states near $E_F$. This can be
particularly important, when there is a strong modulation of the
non-$4f$ background intensity, which is e.g. the case for CeRu$_2$Si$_2$ and
CeNi$_2$Ge$_2$\cite{denlinger:01:0,ehm01}. Here one has the overlap
with Ru~$4d$ and
Ni~$3d$ bands, respectively, that have a comparatively high photoemission cross-section for photon energies in the UV
range.\cite{yeh_lindau:85} In the case of CeCu$_6$, CeCu$_2$Si$_2$, and CeSi$_2$ the
contribution of non $4f$ states in the interesting
energy range is rather small and --- although not negligible ---
without a fine structure on the scale of a few tens of meV. 
Another method to extract the $4f$ spectra from the net photoemission
signal is {\em resonant photoemission spectroscopy} (RESPES) at the Ce
$3d$ or $4d$ absorption edges.\cite{duo:98} Although
this method is limited in energy resolution and therefore not suitable
for investigations of the fine structure on the meV scale, the
relative intensity and energetic
position of the ionization peak ($4f^0$) can be investigated in detail. For all the
systems measured in this work we find values for the $f^0$-energy
(partly from literature \cite{chiaia:97,lawrence:93,ehm_diss}) that
amounts to $1.7$~eV, $2.6$~eV, $2.2$~eV, $2.2$~eV, and $2.35$~eV for
CeCu$_6$, CeCu$_2$Si$_2$, CeRu$_2$Si$_2$, CeNi$_2$Ge$_2$, and CeSi$_2$,
respectively, similar to the values from non-resonant PES.\cite{garnier97}  
As described above, we used these values as benchmarks for the
$\varepsilon_f$ values in the parameter sets of the NCA
calculations (see figure captions 
Figs.~\ref{fig:CeCu2Si2}--\ref{fig:CeSi2}).

\subsection{Quantitative spectral analysis: Fitting with NCA} 

As described in Appendix~\ref{norm_proc} we iteratively fit the
normalized photoemission data for a quantitative data analysis. This
analysis yields {\em one} characteristic model parameter set for each
compound, which is used for the calculation of the spectra at {\em all} investigated
temperatures.  The resulting parameters are given in the captions of
Figs.~\ref{fig:CeCu2Si2}--\ref{fig:CeSi2}

In addition, each parameter set can be assigned to a certain Kondo
temperature $T_K$, which is the Kondo temperature of the respective system.
To get the $T_K$ values we calculated the NCA spectra for each
parameter set in the unitarity limit at $T\!=\!0.1\times T_K$. 
The linewidth of the Kondo resonance in this low-temperature limit defines 
immediately the Kondo temperature by FWHM$\sim k_BT_K$ due to the
scaling properties of the SIAM. The resulting $T_K$'s are given in
Tab.~\ref{tab:TK_CEF} together with the crystal field energies
$\Delta_{CF}$ and the Kondo temperatures determined
from other experimental methods, i.e. INS and transport measurements.
One should note that these reference bulk values are spread
over a wide range, indicating some systematic uncertainties in the determination. 

For all compounds
we find an excellent quantitative agreement of our $T_K$ values with
the Kondo temperatures 
determined by INS
studies.\cite{knopp:88,rossat-mignod:88,bredl:84,horn:81,severing:89:0,galera:89}
In addition, we find a good coincidence of the crystal field
splittings for CeCu$_6$, CeCu$_2$Si$_2$ and CeSi$_2$
with the INS values from Refs.~\onlinecite{goremychkin:93,galera:89}. In the case of
CeRu$_2$Si$_2$ and CeNi$_2$Ge$_2$ the CF energy separations could not
be determined experimentally upto now, mainly
because of uncertainties in the assignment of the CF level scheme. \cite{grier:88,knebel:97}
Therefore in the case of CeRu$_2$Si$_2$ we compare our values with
theoretical results proposed by
Zwicknagl {\em et al.},\cite{zwicknagl:92} based on specific heat
measurements\cite{grier:88} and again we find an excellent agreement.
A recent study on CeNi$_2$Ge$_2$
claims the first observation of CF structures in INS
experiments on this compound, giving values that correspond rather well
with our results.\cite{frost:00}

In summary, the comparison shows that all of our results are in reasonable agreement with the
values from the other methods. This is surprising because of two
possible problems in our method: 1.) In contrast to transport and INS measurements is PES  a
{\em surface sensitive} technique and the surface
properties of a rare-earth compound are not necessarily identical with
the bulk properties, in particular the Kondo temperatures might be
completely different. 2.) The SIAM, on which the NCA
calculations of the spectra are based, contains several significant
simplifications that might influence a quantitative description of
width and position of the Kondo resonance.

\begin{figure*}[tb]
  \begin{center}
    \includegraphics[width =0.55\textwidth,angle=-90]{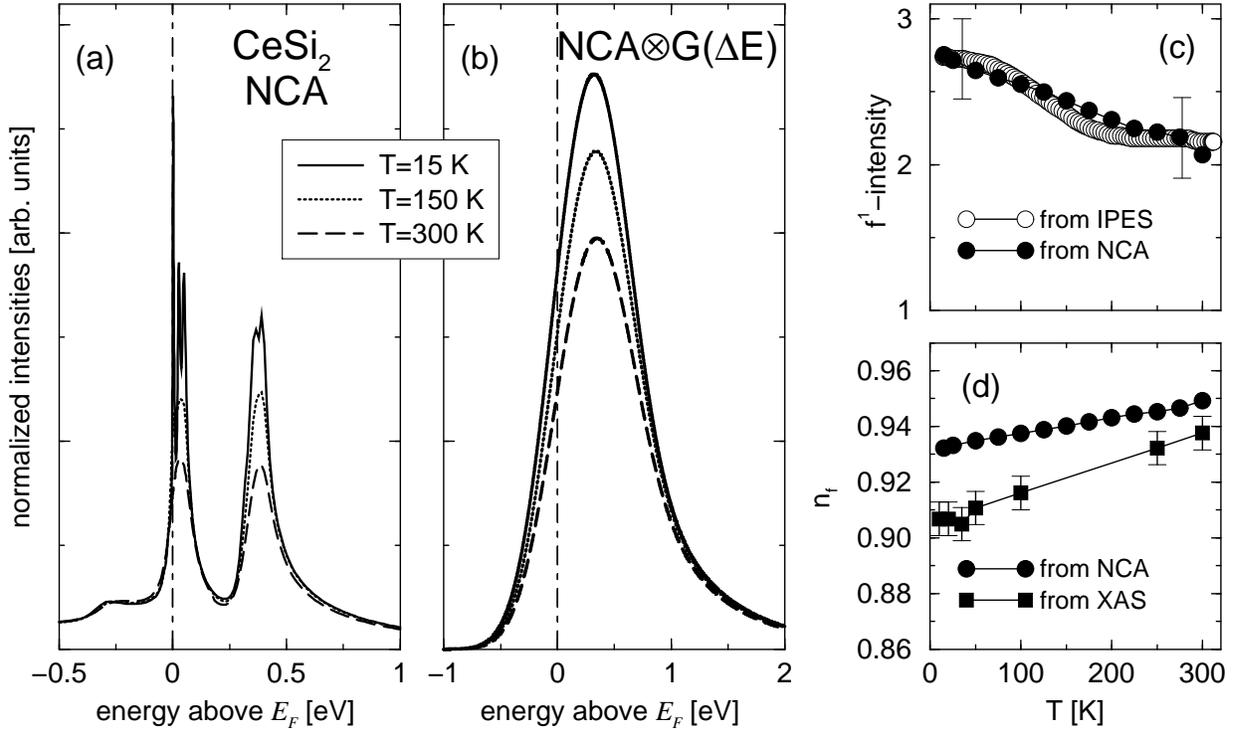}    
    \caption{Left panels: NCA spectra for CeSi$_2$ (parameters given in the
captions of Fig.~\ref{fig:CeSi2}). (a) Raw and (b) convoluted with a Gaussian of
      FWHM$=600$~meV. Panel (c) shows the (normalized) temperature dependence 
      of the $f^1$-intensity as calculated from the convoluted
      NCA-spectra in panel (b), in comparison with the results of
      IPES-studies.\cite{grioni:97} Panel (d)
      shows the temperature dependence of the $4f$ occupation $n_f$ as
      resulting directly from the NCA calculations. These values are
      compared with the ones extracted from XAS measurements at the
      Ce~$L_3$ edge.\cite{grazioli:01}}
    \label{fig:nf_pes_ipes_xas}
  \end{center}
\end{figure*}

Considering the first point, one has to mention that there is an
ongoing debate in how far PES with excitation 
energies in the VUV-range will be able to
reflect the {\em bulk properties} of Ce compounds. 
At these photon energies (and the respective photoelectron kinetic energies
in the range of 40~eV for He~II) the information depth is at a minimum
of a few lattice constants.\cite{huefner} Therefore, one performs additional
photoemission measurements at higher photon energies to increase the
information depth. PES with photon energies in the soft x-ray range,
e.g. in resonance
with the Ce $3d$--$4f$ absorption edge at $h\nu\approx880$~eV, is regarded as
a sufficiently bulk sensitive method with an information depth of about three times
larger than for He~II.\cite{laubschat:90:1} On this
basis Sekiyama {\em et al.}\cite{sekiyama:00:2,sekiyama:00:1,suga:01} have
performed $3d$--$4f$ RESPES studies on CeRu$_2$Si$_2$ with a noticeably
high energy resolution ($\Delta E\!=\!120$~meV). They compared the
experimental spectra with results from
NCA calculations, including CF splittings of $51.6$~meV and $68.8$~meV,\cite{sekiyama:00:2}
and found a good coincidence
between the RESPES data and the NCA spectrum. The significant
difference
between these crystal field energies and our values prompt us to
investigate this point further. We found the following results:
\begin{itemize}
\item[(i)] If we perform NCA calculations with Sekiyama's
parameter set at $T\!=\!20$~K we find clearly separate CF peaks
  below and above $E_F$. As seen in
  Fig.~\ref{fig:CeRu2Si2} we positively cannot observe such structures
  below $E_F$ in our high-resolution low-temperature spectra. In addition,
  these CF energies
  would be too large to reproduce the observed temperature dependence 
  in Fig.~\ref{fig:CeRu2Si2}, right panel.
\item[(ii)] If we convolute the NCA spectra at $T\!=\!20$~K
  with our parameter set (see captions Fig.~\ref{fig:CeRu2Si2}) with a
Gaussian to describe an energy resolution of $\Delta E\!=\!120$~meV, we can reproduce Sekiyama's
  $3d$--$4f$ RESPES spectrum with the same accuracy.
\end{itemize}
Therefore we conclude, that the RESPES results on
CeRu$_2$Si$_2$ do
not contradict our results and, further more, do not indicate a
quantitative deviation of the crystal field energies extracted from
surface sensitive from the bulk values.

There is another possible consistence check of our analysis, using the
comparison with
temperature dependent (resonance) inverse photoemission spectroscopy
(RIPES), which immediately has access to the spectra above the Fermi
level where for Ce-systems the main spectral $4f$ weight of the Kondo
resonance and its satellites appears. Again the energy
resolution of this method is not 
sufficient to resolve the fine-structure of the $4f^1$ features,
namely the Kondo resonance and its crystal
field and spin-orbit satellites ($J\!=\!7/2$). A second broad peak
appears around $U-|\epsilon_f|$ and corresponds to the
two-electron final state $f^2$. However, the NCA spectra, which we get from our fit of the
PES data, contain also the spectral information {\em above} the Fermi
level. For the following comparison we have chosen CeSi$_2$ because there
exist detailed RIPES\cite{grioni:97} and x-ray absorption (XAS)
investigations\cite{grazioli:01} on the temperature dependence of the 
$4f$ occupation number $n_f$, which can be determined from the
intensity ratio of the $f^1$ and $f^2$ features. 

To get this temperature dependence from our results we model the
inverse photoemission spectra by a convolution of the NCA
spectrum (with parameters from fitting the PES data) above the Fermi level (i.e. $s(E,T)\times[1-f(E,T)]$ at the respective
temperature, cf. Appendix~\ref{norm_proc}) with a Gaussian describing
the finite energy resolution of the RIPES experiment. The maximum position
and the line width of the result --- including Kondo resonance, CF satellites and
spin-orbit ($J\!=\!7/2$) partner --- are identical to the experimental
values within the experimental errors.\cite{grioni:97}
The left two panels of Fig.~\ref{fig:nf_pes_ipes_xas} shows (a)
the unbroadened NCA spectra a three different temperatures and (b) the
result of the convolution ($\Delta E\!=\!600$~meV). From this modeled
RIPES spectrum we take the temperature dependent integrated intensity
for a comparison with the experimental temperature dependence.

Panel (c) of Fig.~\ref{fig:nf_pes_ipes_xas} gives the temperature
dependence of the relative integrated RIPES intensities $I(f^1)/I(f^2)$ of
CeSi$_2$ published in Ref.~\onlinecite{grioni:97} (open circles). This intensity
ratio, proportional to $1\!-\!n_f$, clearly increases at low temperatures
and the curve even suggests the existence of the plateaus
well above and below $T_K\!\approx\!30$~K as expected from the theory. 
We compare this behavior with the integrated intensity from
Fig.~\ref{fig:nf_pes_ipes_xas} (b), scaled linearly to match the
RIPES value at the minimum temperature of $15$~K (there is no $f^2$
peak since $U_{ff}=\infty$ in our model). A clear coincidence of
the two results is
obvious, in particular if one takes the error bars of the RIPES
experiment into account. However, the $S$-shape of the experimental
temperature dependence is less pronounced.

From an analysis of the NCA spectra one can also immediately determine
the $4f$ occupation number $n_f$. Experimentally, this $4f$ occupation
can be measured by x-ray absorption spectroscopy (XAS), e.g. at the
Ce~$L_3$ edge at $h\nu\approx5720$~eV. Although this method is truly
bulk sensitive, one has to know that the final state consists of an additional electron in
the $5d$ valence states, which can lead to a slightly modified (final
state) $4f$ occupation sampled by this method. However,
Fig.~\ref{fig:nf_pes_ipes_xas} (d) gives the temperature 
dependence of $n_f$ as extracted from an XAS experiment at the Ce~$L_3$
edge of CeSi$_2$.\cite{grazioli:01} Although the shape of both
curves is very similar, the XAS values are typically $0.01$--$0.03$
smaller than the NCA numbers. The authors in Ref.~\onlinecite{grazioli:01}
explain this difference by surface effects in PES, but this a) can be
ruled out by our previous observations, and b) is usually exceeded by
the final state effect in XAS.\cite{malterre91,lawrence94} However, the
qualitative agreement between XAS and PES/NCA result is reasonable
and indicates that our data analysis is reliable.

Considering the surface effects in the special case of CeSi$_2$, which
has the highest $T_K$ of all Ce systems investigated in this work, one
should note that the published Kondo temperatures, including the
values from spectroscopy methods, differ significantly
in a range of $T_K\!\approx\!22$~K to $140$~K;\cite{mori:84,galera:89,grazioli:01,patthey:87,yashima:82:1}
the given Kondo temperatures of the other compounds have a comparatively small scatter.
This leads to the impression that CeSi$_2$ can be regarded as
borderline system between $\gamma$-Ce- and $\alpha$-Ce like materials, in
which the difference between the hybridization strengths at the
surface (or surface near region)
and the bulk becomes evident.

\section{Conclusions}
In the present paper we have demonstrated that high-resolution
photoemission allows a detailed and quantitative investigation of the
$4f$ spectral features of low $T_K$ cerium compounds close to the Fermi level. The 
Kondo resonance, which for Ce-systems has a maximum above $E_F$, could
be restored for all investigated $\gamma$-Ce like HF systems by
application of a well known normalization procedure. The resulting
$4f$ spectra can be iteratively fitted by NCA calculations based on the SIAM, using
individual model parameter sets including the spin-orbit and, in
particular, the crystal field one-electron energies
$\epsilon_f$. Our results demonstrate that the consideration of the
crystal-field splitting and the corresponding fine structure in the
$4f$ spectra is of high importance for the consistency of the
photoemission results with the thermodynamic properties. Surface
effects can be ruled out for the investigated $\gamma$-Ce like
systems. In spite of the known limitations of the model, our data analysis yields
Kondo temperatures and crystal field energies that are in
surprisingly good agreement with values from other experimental
methods published in the literature. For the
investigated temperature range, the description by a local model ---
namely the SIAM --- is obviously sufficient to describe the
photoemission data with an energy resolution of the order of the Kondo
temperature. 

\begin{acknowledgments}
This work was supported by the Deutsche Forschungsgemeinschaft, grant
nos. Hu~149/19-1 and Re~1469/4-3 (D.E., S.H., F.R.),  
through SFB 608 (J.K.) and by the Virtual Institute for Research on 
Quantum Phase Transitions at the University of Karlsruhe (P.W.,
H.v.L.). Particularly one of the authors
(S.H.) thanks O. Gunnarsson for many enlightening discussions and
communications over a number of years.
\end{acknowledgments}

\begin{appendix}
\section{Normalization Procedure}
\label{norm_proc}

On the example of the $3d$ bands of Ni(111),
Greber \cite{greber97} has shown that a careful analysis
of photoemission data close to the Fermi level allows an extended access to the
thermally excited spectral features up to
$5k_BT$ above $E_F$. Two conditions for this method have to be fulfilled: 1.) a satellite free and
highly intense photon source must provide low noise data, and 2.) the
total experimental energy resolution $\Delta E$ must be at least
comparable to the thermal broadening $4k_BT$ of the Fermi edge; we meet these two
conditions by using the experimental setup described above. The normalization
method has been established for the qualitative spectroscopic investigation of the
near-$E_F$ range for various compounds, both for angular integrated
and angular resolved data (see
e.g.\ Refs.\ \onlinecite{greber:97,kumigashira:99,pillo:00,forster04})

The experimental photoemission spectrum $I$ can be described as the
density of electronic states (or spectral function) $s(E,T)$
multiplied by the temperature dependent Fermi-Dirac distribution (FDD) $f(E,T)$. To consider
the finite energy resolution, the product must be
convoluted by the spectrometer function $g(E,\Delta E)$, which is
usually approximated by a Gaussian with the full width at half maximum
(FWHM) $\Delta E$. Note that in our case the Gaussian describes nearly
perfectly all reference spectra \cite{nicolay00,reinert_icess8}, independent from the individual
spectrometer settings. Thus one gets the photoelectron spectrum
\begin{eqnarray}
\label{eqn:PES_int}
  I(E,T)&=&\left[s(E,T)\cdot f(E,T)\right]\otimes g(E,\Delta E), 
\end{eqnarray}
plus statistical noise, which we neglect in the following
discussion (see Ref.~\onlinecite{greber97} for more).
In principle, this relation is also valid for angular resolved spectra, but here the
finite angular resolution must be taken into account too.

The normalization procedure is
realized by dividing the measured spectrum $I(E,T)$ by a Gaussian-broadened
FDD:
\begin{eqnarray}
\label{eqn:normalization}
\displaystyle{\frac{\left[s(E,T)\cdot f(E,T)\right]\otimes g(E,\Delta E)}{f(E,T)\otimes g(E,\Delta E)}}. 
\end{eqnarray}
Because the convolution is not
distributive (i.e. $[s\cdot f]\otimes g \neq [s\otimes g]\cdot
[f\otimes g]$), the result of eq.~(\ref{eqn:normalization}) is {\em not}
equal to the broadened spectral function $s(E,T)\otimes g(\Delta E,E)$. In other words, the
normalization procedure does not {\em a priori} reveal the
intrinsic spectral function. As long as the energy broadening $\Delta
E$ is small compared to the thermal broadening $4k_BT$, the deviation
between the normalization result and $s\otimes g$ is small and the
result gives a useful
qualitative information about the structures in the energy range up to
$5k_BT$ above $E_F$.

However, the normalization can be used for even a {\em quantitative}
analysis of the intrinsic spectral function when an iterative
procedure is applied.\cite{ehm_sces02}

\begin{figure}[tb]
  \begin{center}
    \includegraphics[width = 8.2cm]{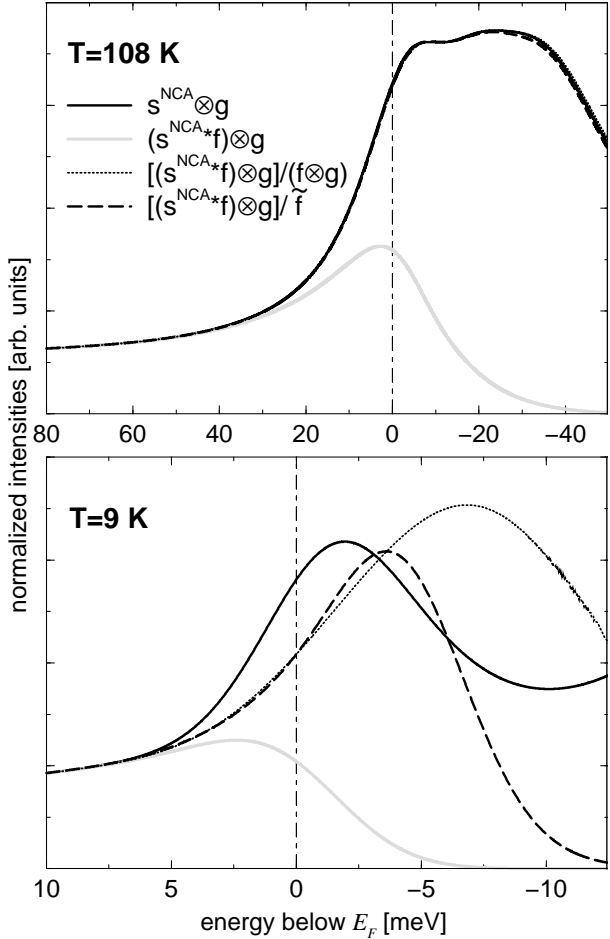}    
    \caption{Application of the normalization procedure of
      eq.~(\ref{eqn:normalization}) at two different
      temperatures. Top panel: $T\!=\!108$~K$\gg \Delta E/k_B$, lower
panel: 9~K$\ll\Delta E/k_B$. The energy resolution was $\Delta
E\!=\!5.4$~meV. The grey curves give the modeled photoemission spectra,
including FDD and Gaussian broadening, the black solid lines represent
the (broadened) intrinsic spectral functions.} 
    \label{fig:normalization}
  \end{center}
\end{figure}

The influence of this effect can be
demonstrated by its application to model spectral functions
(similar to the theoretical NCA spectra with realistic parameters) at
different temperatures. The top panel of Fig.~\ref{fig:normalization}
gives the broadened intrinsic spectrum $s^{NCA}\otimes g$ with $\Delta
E\!=\!5.4$~meV (black solid line) and the modeled photoelectron spectrum $\left
[s^{NCA}\!\cdot\!f\right]\otimes g$ according to  
eq.~(\ref{eqn:normalization}) (grey line). We divide the latter by
$f\otimes g$ with $T\!=\!108$~K and get the dotted curve, that differs
only little from the desired result $s^{NCA}\otimes g$ over the
investigated temperature range of approximately up to $5k_BT$. 

In this high-temperature case
one can replace --- as also stated in Ref.~\onlinecite{greber97} ---
the denominator in
eq.~(\ref{eqn:normalization}), i.e. the broadened FDD $f\otimes g$, by
a bare FDD $\tilde{f}(E,T_{eff})$ with an effective temperature of
$T_{eff}=\sqrt{T^2+(\Delta E/4k_{\rm B})^2}$. As shown in the top
panel of Fig.~\ref{fig:normalization} this simplified
normalization is in good agreement with $s^{NCA}\otimes g$.

For low temperatures $T\lesssim \Delta E/k_B$, the difference between
the normalized spectrum --- with convoluted or effective FDD ---
and the intrinsic spectrum becomes significant (see lower panel of
Fig.~\ref{fig:normalization}). Both the resulting width and the 
maximum position deviate considerably from the right numbers. 
Surprisingly, the shape of the the normalization with $\tilde{f}$ is
closer to the intrinsic spectrum $s\otimes g$ in the present case than
the one with $f\otimes g$. As described above, the
difference between the individual curves decreases with increasing temperature, finally matching
when $T>>\Delta E$. 

However, at low temperatures the spectral shape of the normalized
spectrum must not be used to
determine line width and position directly. Instead one has to {\em
fit}
the spectra iteratively by comparing (e.g. by calculating $\chi^2$) the normalized experimental data
with a theoretical spectrum treated the same way. This means one has
to model the experimental spectrum according to eq.~\ref{eqn:PES_int},
and use the same normalization procedure for this modeled spectrum
and for the experimental data. If the curves match, one can get the
position and the line width from the bare $s^{NCA}(E,T)$ included in
the modeled function.
\end{appendix}



\end{document}